# Growth and atomically resolved polarization mapping of ferroelectric Bi$_2$WO$_6$ thin film

Jihwan Jeong[1,2,‡], Junsik Mun[3,1,‡], Saikat Das[1,2,*], Jinkwon Kim[1,2], Jeong Rae Kim[1,2], Wei Peng[1,2], Miyoung Kim[3,1], and Tae Won Noh[1,2]

[1] Center for Correlated Electron Systems, Institute for Basic Science (IBS), Seoul 08826, Korea.

[2] Department of Physics and Astronomy, Seoul National University, Seoul 08826, Korea.

[3] Department of Materials Science and Engineering and Research Institute of Advanced Materials, Seoul National University, Seoul 08826, Republic of Korea.

**ABSTRACT**

Aurivillius ferroelectric $Bi_2WO_6$ (BWO) encompasses a broad range of functionalities, including robust fatigue-free ferroelectricity, high photocatalytic activity, and ionic conductivity. Despite these promising characteristics, an in-depth study on the growth of BWO thin films and ferroelectric characterization, especially at the atomic scale, is still lacking. Here, we report pulsed laser deposition (PLD) of BWO thin films on (001) $SrTiO_3$ substrates and characterization of ferroelectricity using the scanning transmission electron microscopy (STEM) and piezoresponse force microscopy (PFM) techniques. We show that the background oxygen gas pressure used during PLD growth mainly determines the phase stability of BWO films, whereas the influence of growth temperature is comparatively minor. Atomically resolved STEM study of a fully strained BWO film revealed collective in-plane polar off-centering displacement of W atoms. We estimated the spontaneous polarization value based on polar displacement mapping to be about $54 \pm 4$ μC cm$^{-2}$, which is in good agreement with the bulk polarization value. Furthermore, we found that pristine film is composed of type-I and type-II domains, with mutually orthogonal polar axes. Complementary PFM measurements further elucidated that the coexisting type-I and type-II domains formed a multidomain state that consisted of 90° domain walls (DWs) alongside multiple head-to-head and tail-to-tail 180° DWs. Application of an electrical bias led to in-plane 180° polarization switching and 90° polarization rotation, highlighting a unique aspect of domain switching, which is immune to substrate-induced strain.

**INTRODUCTION**

$Bi_2WO_6$ (BWO), in which a perovskite-like $WO_6$ block alternates between two $Bi_2O_2$ sheets, has the simplest structure among Aurivillius compounds[1,2]. In bulk, BWO undergoes a paraelectric-ferroelectric phase transition below 950°C and develops a polar orthorhombic phase with a large spontaneous polarization of about 50 μC cm$^{-2}$. The high Curie temperature and large polarization, combined with the generic fatigue-free[3,4], low-leaky behaviors of Aurivillius compounds[5], make BWO an attractive candidate for ferroelectric applications. Also, a low-energy barrier for electrical switching of ferroelastic domains, which involves a 90° polarization rotation, enables reversible control of in-plane strain in BWO thin films, which may be relevant for electromechanical and magnetoelectric device applications[6]. Besides its core appeal in ferroelectric applications, BWO is known for its excellent photocatalytic activity and ionic conductivity[7,8]. A recent theoretical study suggested that the band structure of BWO is Rashba spin-split[9]. Switching of the ferroelectric polarization, moreover, is predicted to provide a means to control the Rashba spin-splitting electrically, thereby opening up novel spintronics functionalities[10].

To fully utilize the functionalities of BWO, an understanding of BWO thin-film growth and the microscopic origin of ferroelectricity is needed. It is believed that the ferroelectricity in BWO is linked to two distinct polar distortions[2,6]; one involves the displacement of tungsten (W) atoms against the oxygen octahedra, and the other is the result of the relative shift between the $Bi_2O_2$ layers and the $WO_6$ blocks. Although atomic-scale structural characterization of BWO films has been reported[6], atomically resolved polar displacement mapping, which is critical to understanding the nature of ferroelectricity, has not. Pulsed laser deposition (PLD) growth of BWO thin films and piezoresponse force microscopy (PFM) characterization of ferroelectricity, meanwhile, have been reported separately, but there has been no systematic study that aimed to construct the growth phase diagram for BWO. Moreover, electrical switching of ferroelectric domains has been investigated using fully or partially relaxed orthorhombic BWO thin films with lattice parameters similar to the bulk value[6]. The corresponding ferroelectric domain configuration and its response to an electric stimulus in a coherently strained film have remained unexplored.

Here, we first performed a parametric study of the PLD growth of BWO thin films as a function of the growth temperature ($T_{sub}$) and background oxygen partial pressure ($P_{O_2}$), enabling the construction of a growth-phase diagram. Subsequently, we investigated the microstructure and performed atomically resolved polar displacement mapping of a coherently strained BWO film using scanning transmission electron microscopy (STEM). By combining high-angle annular dark field (HAADF) and annular bright field (ABF) STEM imaging, we revealed the coexistence of type-I and type-II domains that are distinguishable by subtle distortions within $Bi_2O_2$ blocks. Atomically resolved HAADF-STEM imaging allowed direct mapping of the polar displacement of W atoms, which yielded, on average, an in-plane polarization of about $54 \pm 4$ μC cm$^{-2}$. Furthermore, polarization orientation within type-I and type-II domains are mutually orthogonal, forming a 90° domain wall (DW). We corroborated this observation using PFM imaging and demonstrated full electrical control of ferroelectricity, including 90° polarization rotation.

**RESULTS AND DISCUSSION**

To construct a growth-phase diagram of BWO thin films, we grew a series of samples on (001)-oriented $SrTiO_3$ (STO) substrates using the PLD technique. The films were grown by varying the oxygen partial pressure ($P_{O_2}$) and $T_{sub}$, while other PLD parameters were kept fixed (see Materials and Methods). Figure 1a shows the results of symmetric $2\theta$-$\omega$ X-ray diffraction (XRD) scans performed on BWO films grown at a substrate temperature of $T_{sub} = 600$°C by varying $P_{O_2}$ from 5 mTorr to 250 mTorr. For $P_{O_2} \leq 50$ mTorr, {00$l$} BWO peaks are not discernable, only weak peaks at $2\theta = 23.5°$ and $29.55$–$29.8°$ can be observed, which may be associated with W-rich impurity phases[13–15]. In the intermediate range from $P_{O_2}$ = 50 mTorr to 75 mTorr, {00$l$} peaks coexist with a broad feature centered around $2\theta$ ~ 28.3–28.5°, which closely proximates to the bulk (113) peak position. Increasing the pressure further to $P_{O_2} \geq$ 125 mTorr led to the stabilization of the pure (001)-oriented BWO phase. Therefore, $P_{O_2}$ = 50–75 mTorr marks the transition from growth instability to the stable single-crystalline (001) phase through an intermediate mixed-phase state. Notably, among the films grown under high pressure, the one grew at $P_{O_2}$ = 125 mTorr exhibits the most pronounced (006) Bragg peak, alongside Kiessig fringes covering an extended $2\theta$ range, thus implying high crystallinity. Based on these Kiessig fringes, we estimated the thickness of the film to be around 20 nm (~12 unit cells).

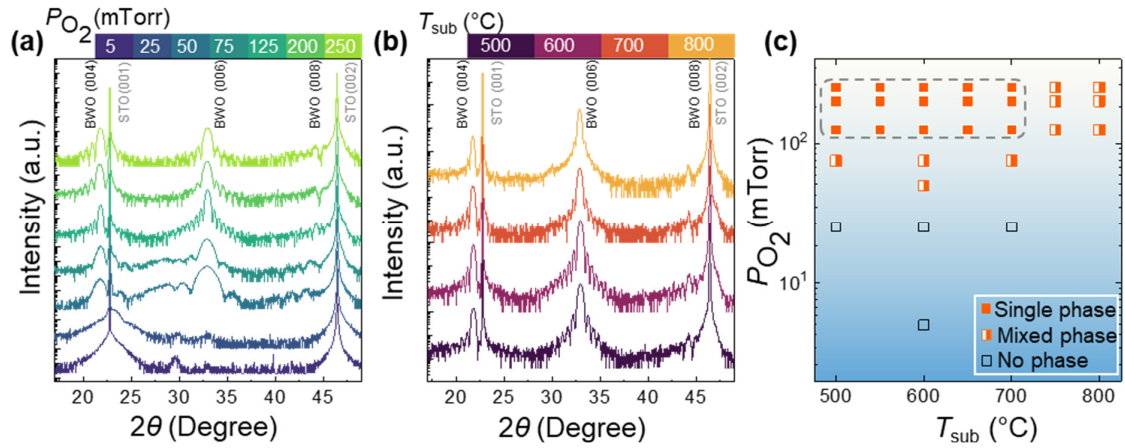

**Figure 1.** Structural characterization and growth-phase diagram of BWO thin films. (a) XRD patterns of BWO thin films grown at $T_{sub}$ = 600°C, while varying $P_{O_2}$ from 5 mTorr to 250 mTorr. (b) XRD patterns of BWO thin films grown at $P_{O_2}$ = 125 mTorr, by varying $T_{sub}$ from 500°C to 800°C. For clarity, the XRD spectra in Figs. (a) and (b) are vertically offset, and spectra spanning the $2\theta$ values in the range 17–49°, which covers the most intense BWO (006) Bragg peak, are shown. (c) Growth-phase diagram constructed from the XRD study of BWO films as a function of $T_{sub}$ and $P_{O_2}$. The dashed gray rectangle marks the optimal growth window for the (001)-oriented BWO thin film.

Next, to determine the impact of temperature on BWO film growth, we varied $T_{sub}$ from 500°C to 800°C in steps of 50°C, while $P_{O_2}$ was set to 125 mTorr. Figure 1b shows representative XRD spectra of films grown at $T_{sub}$ = 500°C, 600°C, 700°C, and 800°C. Compared to the background $P_{O_2}$, the influence of $T_{sub}$ on BWO growth is moderate. The epitaxial (001)-oriented phase can be stabilized effectively at all $T_{sub}$, with the emergence of a weak secondary phase (2θ ~ 30°) when $T_{sub} \geq$ 750°C. Also, as evident from Figure 1b, with increasing $T_{sub}$, the Kiessig fringes disappear, implying a decrease in crystalline quality. A systematic rocking curve analysis (Figure S1) further corroborates this conjecture and also confirms that the BWO film with the highest crystallinity can be grown at $T_{sub}$ = 600°C and $P_{O_2}$ = 125 mTorr. For this film, the full width at half maximum (FWHM) of the (006) peak is ~ 0.04°, an order of magnitude lower than previously reported values[11,12].

Based on this parametric study, we constructed a growth-phase diagram for BWO films as a function of $T_{sub}$ and $P_{O_2}$ (Figure 1c). Figure 1c indicates that while the epitaxial (001) BWO phase could be stabilized over a wide temperature range, the phase stability strongly dependent on the oxygen partial pressure. We identified the optimal window for growing the pure (001)-oriented BWO phase (dashed rectangle in Figure 1c) from this growth-phase diagram. Systematic atomic force microscopy (AFM) measurements reveal that films from the optimal growth window exhibit a step-terrace structure. However, well-defined steps and low surface roughness (Figure S2) can be obtained only for $T_{sub} \leq$ 600°C. Overall, the rocking curve and AFM topography analyses indicate that the growth parameters, $T_{sub}$ = 600°C and $P_{O_2}$ = 125 mTorr yield (001)-oriented BWO films with the highest crystallinity. In the remainder of this paper, we focus on BWO films grown using these optimized parameters.

To assess the strain state of the optimized BWO films, based on the XRD study, we estimated the $c$-axis lattice parameter of the BWO film to be around 16.301 ± 3.2E-4 Å. This value is about 0.79% smaller than the bulk value of ~ 16. 430 Å[16]. Such contraction of the *c*-axis lattice parameter is to be expected, as the underlying STO (*a* = 3.905 Å) imposes, on average, an in-plane tensile strain of about + 1.3% on BWO (*a* = 5.456 Å, *b* = 5.436 Å). Reciprocal space mapping (Figure S3) further confirmed that the film was under tensile strain, adopting in-plane lattice parameters *a* = 5.5205 ± 1.6E-4 Å and *b* = 5.5167 ± 1.9E-4 Å. These values are fairly comparable and agree closely with $\sqrt{2}a_{STO}$ = 5.5225 Å, implying that the film is fully strained.

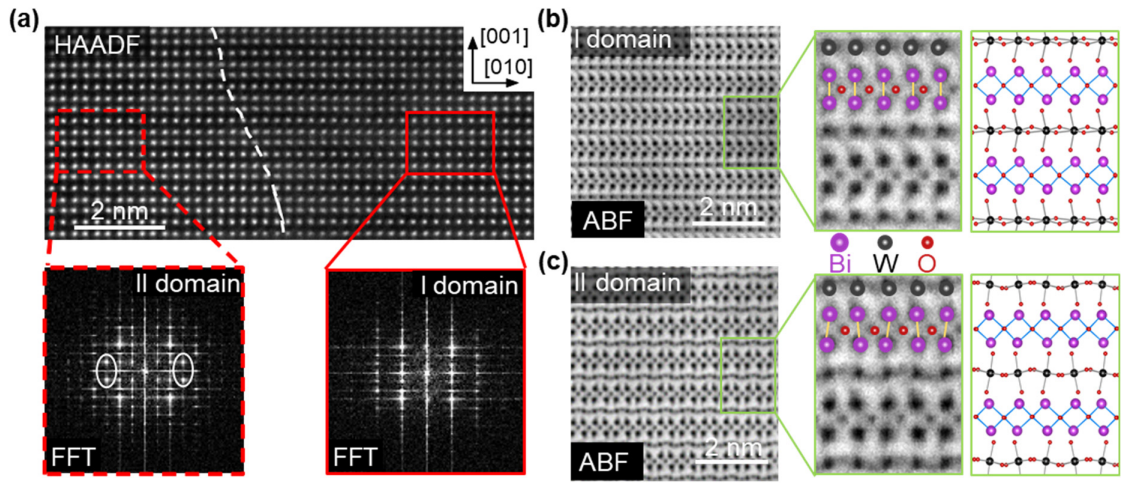

**Figure 2**. Structural characterization of BWO films by STEM. (a) High-resolution HAADF-STEM image along the BWO [100] zone axis (top panel). The bottom panel shows FFT images of the selected regions, marked by the solid and dashed rectangles, which we identified as the type-I and type-II domains. The dashed line in (a) is a guide for the eye and marks the boundary between the two domains. (b) (c) ABF-STEM images of the type-I and type-II domains and the corresponding crystal structure models. The magnified parts of the ABF-STEM images are also shown, highlighting that the Bi-Bi distance was fairly constant along the BWO [010] axis within the type-I domain but alternated between 'long' and 'short' values within the type-II domain.

The low-magnification STEM measurements (Figure S4) reveal that the optimized BWO film is coherently grown over a wide lateral scale. Although free of excessive defects, the out-of-phase boundary typical of layered materials can be observed emanating from a defective region near the interface with the substrate. In contrast, the BWO film grown at $T_{sub}$ = 600°C and $P_{O_2}$ = 75 mTorr exhibits excessive crystalline defects and contains a bismuth (Bi)-deficient nonlayered phase (Figure S5). This observation is consistent with the XRD analysis, and further confirms that the oxygen partial pressure dominates the phase stability of BWO.

Figure 2a shows a representative high-resolution HAADF-STEM image of the optimized BWO film along the BWO [010] (or STO [$\bar{1}$10]) zone axis[6] (see Materials and Methods for details of the measurements). Fast Fourier transformation (FFT) of the HAADF-STEM image (bottom panel of Figure 2a) reveals the presence of two distinct structural domains, which we refer to hereafter as the type-I and type-II domains. The FFT image of the type-II domain exhibits half-order spots (marked by ellipses), which are absent in the FFT image of the type-I domain. These half-order spots typically suggest doubling of the lattice periodicity.

To elucidate the structural dissimilarity between the type-I and type-II domains and the origin of these half-order spots, in Figure 2b and 2c, we show the corresponding ABF-STEM images and the schematic structural models. Unlike HAADF-STEM imaging, which is sensitive only to the heavier W and Bi atoms, the ABF mode can also probe the lighter oxygen (O) atoms (red spheres in Figure 2b,c), providing complete structural information.[17,18] The ABF-STEM images, especially the magnified images, show that the Bi-Bi distance along the BWO [010] (or STO [$\bar{1}$10]) direction is fairly constant in the type-I domain (Figure 2b). However, alternating "short" and "long" values are observed in the type-II domain (Figure 2c). In addition, the zig-zag pattern of the oxygen network in the basal plane of the WO$_2$ layer is discernable in Figure 2c. These features can be well accounted for by the structural model for the polar orthorhombic phase (space group $P2_1ab$) projected along the crystallographic *a*-axis (Figure 2c). Meanwhile, the ABF image from the type-I domain is well accounted for by the orthorhombic structure projected along the crystallographic *b*-axis (Figure 2b). This implies that the type-I and type-II domains are structural twins, where the *a*/*b* axis rotates in-plane by 90° across the boundary marked by the dashed line in Figure 2(a). The alternating "short" and "long" Bi-Bi distance and the "zig-zag" distortion of the oxygen network observed in the type-II domain effectively doubles the in-plane lattice periodicity. The former contribution dominates the HAADF signal, thereby accounting for the half-order spots in the FFT image in Figure 2(a). Notably, the observation of coexisting domains is consistent with a previous study[6], in which the lattice constants of these domains were found to adopt bulk values of 5.46 Å and 5.43 Å. Therefore, it was argued that the boundary separating these domains accommodated about 0.4% ferroelastic strain. According to the XRD study, however, the in-plane lattice parameters are comparable. Thus, we expect negligible ferroelastic strain across the domain walls separating the type-I and type-II domains.

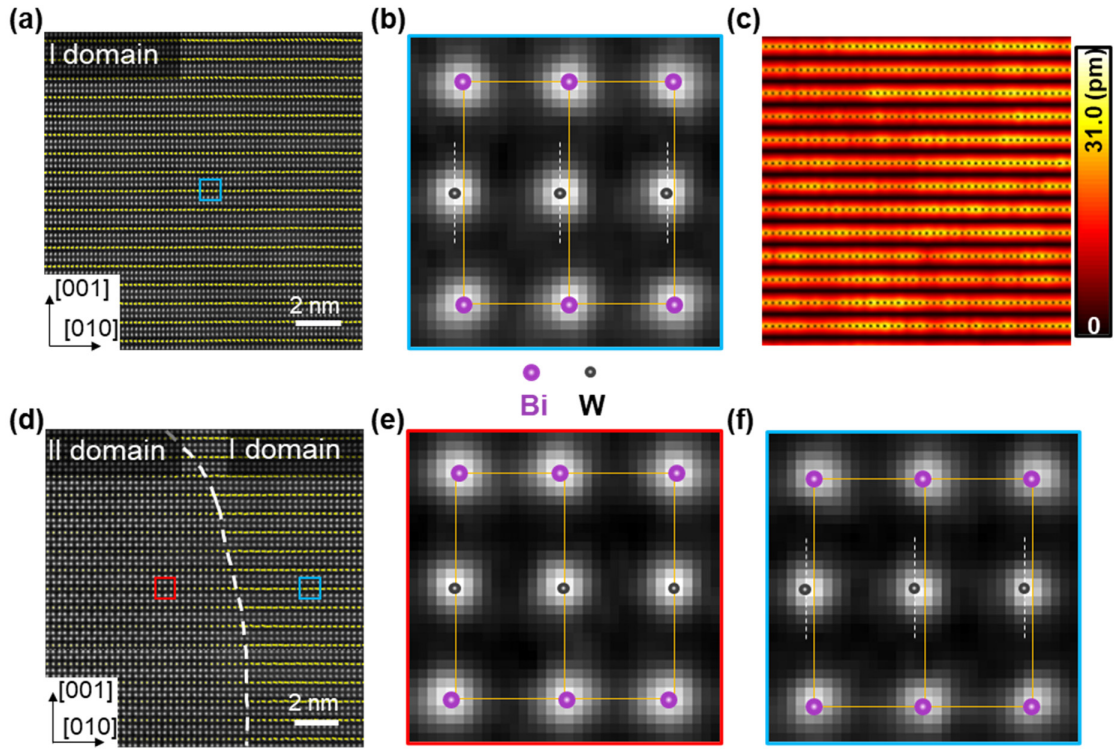

**Figure 3**. Atomic-scale characterization of polar displacement. (a) High-resolution HAADF image taken within the type-I domain. (b) Magnified HAADF image showing the polar displacement of W atoms (black spheres) relative to the Bi atoms (magenta spheres) within the area marked by the cyan-colored rectangle in (a). (c) False color maps showing the layer-resolved magnitude of polar displacement within the type-I domain. (d) High-resolution HAADF image across the boundary between the type-I and type-II domains. The dashed line is a guide for eyes and marks the boundary between the two domains. (e), (f) HAADF images magnifying the regions marked as red and cyan-colored rectangles in (d), showing the non-polar (e) and polar arrangement of W atoms (f). The yellow arrows in (a) and (d) mark the direction of polar displacement. These HAADF images were obtained along the BWO [100] zone axis.

Next, we focus on the atomic-scale characterization of the polar displacement. Using high-resolution HAADF-STEM measurements, we estimated the polar off-centering displacement of W atoms relative to the nearest Bi atoms of the $Bi_2O_2$ block. In Aurivillius ferroelectrics containing A-site cations, these nearest Bi atoms are colinear to the A-site cations, which has previously been used to map the polar displacement of B-site atoms[19]. The HAADF-STEM images taken within the type-I domain and across the boundary between the type-I and type-II domains are shown in Figures 3a and 3d, respectively. Within the type-I domain (Figure 3a), the W-atoms were displaced along the BWO $[0\bar{1}0]$ or STO $[1\bar{1}0]$ direction, as indicated by the yellow arrows. In Figure 3(b), a selected part of Figure 3a is magnified; the vertical dashed lines mark the positions of W atoms that were off-centered relative to the Bi columns. The layer-resolved magnitude of this polar displacement is shown in Figure 3c. The magnitude of the polar displacement varied within a layer, with an average value of $D \sim 21.5 \pm 1.5$ pm. Based on the empirical formula $P_S = 2.5\ D$ that relates the polar displacement ($D$ in units of pm) of B-site cations to the spontaneous polarization ($P_S$ in units of μC cm$^{-2}$) in displacive ferroelectrics[20,21], we estimated the average $P_S$ value in BWO to be ~ 54 ± 4 μC cm$^{-2}$. This value is consistent with theoretical calculations and the measured polarization value in bulk single crystals[9,22]. Therefore, the agreement between the bulk polarization value and our result indicates robust ferroelectricity in the BWO film. The robust bulk-like polarization in BWO film may arise from substrate-induced large tensile strain (~ 1.3%), which generally favors in-plane ferroelectricity in oxide ferroelectrics[19,20].

In contrast to the type-I domain, we could not detect the polar displacement of W atoms within the type-II domain, as evident from the HAADF image taken across the boundary between the type-I and type-II domains (Figure 3d). This observation is understandable because, in the polar orthorhombic phase of BWO, polar displacement is along the crystallographic *a*-axis, which for the type-II domain is parallel to the zone axis, inhibiting visualization of the polar displacement. Meanwhile, within the neighboring type-I domain, where the crystallographic *a*-axis is orthogonal to the zone axis, the W atoms are found to be displaced along the BWO [0$\bar{1}$0] direction (Figure 3f), similar to Figure 3b. We also found a region where the displacement of W atoms is along the BWO [010] direction within the type-I domain side of the boundary (Figure S6). Additional STEM images that were taken along the BWO [010] zone axis (Figure S7) further confirm the presence of type-I and type-II domains, albeit the polar displacement within the former domains is along the BWO <100> directions. Therefore, the boundary between type-I and type-II domains (marked by the white dashed line in Figure 3d) marks a 90° DW. Notably, we did not discern any out-of-plane (*i.e.,* parallel to BWO [001] axis) component of the polar displacement around the 90° DW. This implies that polarization rotation is strictly confined to the BWO (001) plane. Besides 90° DW, we found 180° DW, which surprisingly coincided with the out-of-phase boundary (Figure S4).

Overall, the STEM study unambiguously demonstrates that the in-plane ferroelectricity in BWO is linked to polar off-centering of the W atoms, consistent with previous theoretical and structural refinement studies[2,16]. Additionally, we found that two domains coexist in the BWO film, wherein the polar axes are mutually orthogonal, forming 90° DWs. To complement the atomic-scale polarization characterization with micro-scale domain imaging and study their response to electrical bias, we performed PFM measurements. For this purpose, we used a fully strained 11-unit-cell-thick BWO film grown on an 8-nm-thick metallic SrRuO$_3$ (SRO) bottom electrode (see Materials and Methods and Figure S8 and Figure S9).

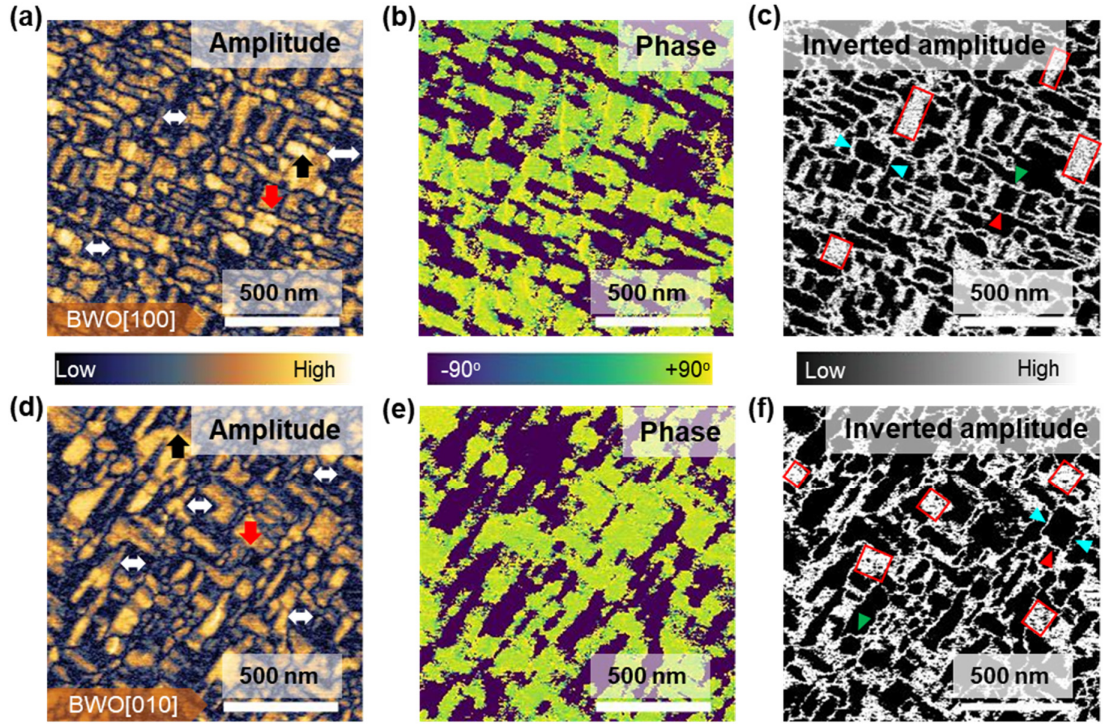

**Figure 4**. In-plane PFM imaging of ferroelectric domains. (a, b, c) PFM amplitude (a), PFM phase (b), and a PFM amplitude image shown using an inverted color scale (c). These images were obtained with the cantilever axis aligned parallel to the BWO [100] direction. (d, e, f), PFM amplitude (d), PFM phase (e), and the PFM amplitude image shown using a reverse color scale (f). The PFM images were obtained with the cantilever axis aligned parallel to the BWO [010] direction. The orientation of ferroelectric polarization within the domains is marked by red, blue, and white arrows in Figure 4 a,d. In Figure 4c,f the representative 90° DWs are marked by red rectangles; the head-to-head, tail-to-tail, and charge-neutral 180° DWs are marked by red, green, and cyan-colored triangles, respectively.

Consistent with the STEM analysis, which showed that the polarization was strictly in-plane, we did not detect any out-of-plane PFM signal in the BWO film. Therefore, the following discussion is restricted to in-plane PFM measurements.

The in-plane PFM images captured by aligning the cantilever axis along the BWO [100] (Figure 4a,b) and [010] (Figure 4d,e) directions show complex pristine domain arrangements in the film. Ferroelectric domains with high PFM amplitude but opposite PFM phases (marked by vertical arrows), alongside domains with negligible PFM amplitude (marked by horizontal arrows) appearing as dark patches in Figure 4a,d can be observed. Because only the polarization components orthogonal to the cantilever axis contribute to the in-plane PFM signal[21], the brighter regions in Figure 4a (Figure 4d) could be assigned to domains with polarization parallel/antiparallel to the BWO [010] and [100] directions. Meanwhile, the darker regions could be assigned to domains with the polarization along the cantilever axis. For better visualization of these domains and DWs, in Figure 4c and 4f, we plot the PFM amplitudes from Figure 4a and 4d using an inverted color scale. Here, the domains with polarization along the cantilever axis and 180° DWs appear in brighter contrast. Therefore, the boundary of these brighter regions (marked by red-colored rectangles) corresponds to the 90° DWs; meanwhile, the bright thin lines are the 180° DWs. Comparing the PFM amplitude and phase data, we can identify charge-neutral, head-to-head, and tail-to-tail 180° DWs, marked by cyan, red, and green-colored triangles, respectively, in Figures 4c and 4f. Despite the presence of head-to-head and tail-to-tail 180° DWs, electrical conductivity was not detected across these DWs.

Finally, to test the electrical switchability of the ferroelectric domains, we poled the BWO film using a trailing electrical field applied through a biased scanning probe tip[22]. Figure 5 shows in-plane PFM images obtained after domain writing with a tip bias of ± 6.5 V, whereby the slow scan axis was set to the BWO [100] (Figure 5a,b) and BWO [010] (Figure 5c,d) directions. For both orientations, the application of a tip bias of -6.5 V (+6.5 V) completely erased the original multidomain state, yielding a monodomain state with the polarization orientation parallel (antiparallel) to the slow scan direction. Note that there is an amplitude contrast between regions poled with -6.5 V and +6.5 V, which gradually diminishes over time (Figure S10), implying that charge injection from the biased tip during poling introduced an additional contribution to the PFM images[23]. Nevertheless, Figure 5 demonstrates that the trailing electric field enabled full in-plane 180° switching and 90° rotation of polarization along four equivalent BWO {100} directions. By studying the bias-dependent switching, we further confirmed that both 180° and 90° switching occurred concurrently in our film (Figure S11). Given that the BWO film was fully strained, we conclude that epitaxial strain has a negligible influence on in-plane ferroelectric switching. The corresponding energy barrier is intrinsically low in BWO, as proposed in an earlier study[6].

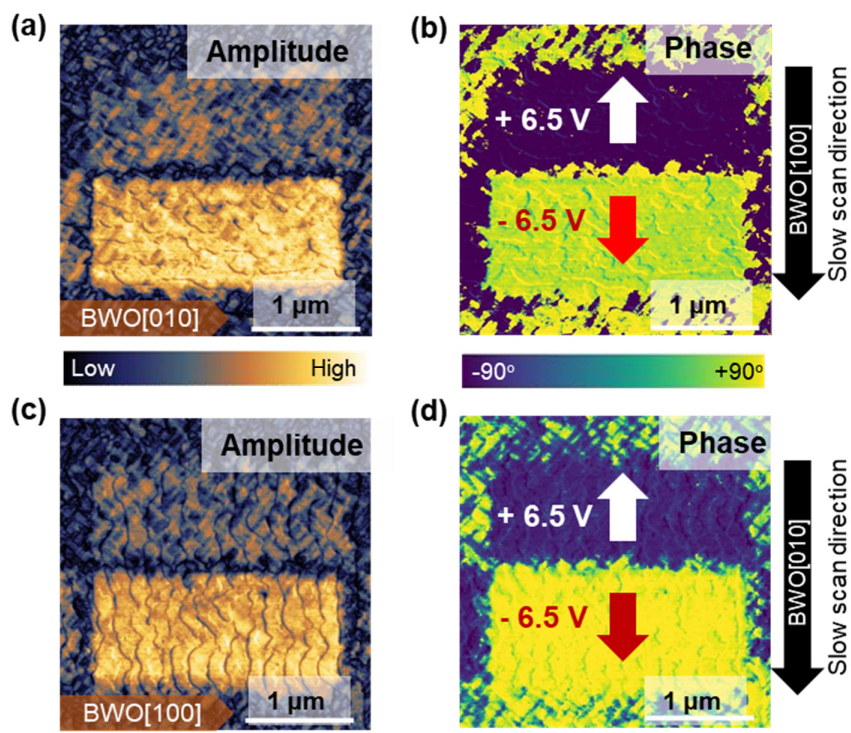

**Figure 5**. Electrical switching of ferroelectric domains. (a, b) PFM amplitude (a) and PFM phase (b) images taken after poling with tip biases of ± 6.5 V. During the electrical poling, the slow (fast) scan axis was set parallel to the BWO [100] ([010]) direction, PFM images are obtained with the cantilever axis aligned parallel to the BWO [010] direction. The red and white arrows indicate IP ferroelectric domains. (c, d) PFM amplitude (c) and PFM phase (d) images after poling with tip biases of ± 6.5 V. During the electrical poling the slow (fast) scan axis was set parallel to the BWO [010] ([100]) direction, PFM images are obtained with the cantilever axis aligned parallel to the BWO [100] direction. The red and white arrows mark the in-plane polarization orientations.

**CONCLUSIONS**

In summary, we comprehensively studied the influence of $P_{O_2}$ and $T_{sub}$ on the PLD growth of (001)-oriented BWO thin films, which highlighted that $P_{O_2}$ critically influences the phase stability of BWO. Meanwhile, we demonstrated that single-phase BWO films can be grown over a relatively wide $T_{sub}$ range. For optimal $P_{O_2}$ and $T_{sub}$ values, fully strained high-quality films with an FWHM value as low as 0.04° can be grown. Using atomically resolved STEM imaging, we unambiguously demonstrated that the ferroelectricity in BWO stems from collective in-plane off-centering displacement of W atoms. We estimated the spontaneous polarization to be about $54 \pm 4$ μC cm$^{-2}$, which is in good agreement with the bulk polarization value. Furthermore, we showed that the BWO film consists of ferroelectric type-I and type-II domains, forming 90° DWs in the pristine state. Complementary PFM characterization further showed that the pristine ferroelectric phase exhibited a multidomain state, in which 90° DWs coexist alongside head-to-head and tail-to-tail 180° DWs. Application of an electrical bias enables full in-plane 180° switching and 90° polarization rotation, highlighting a unique feature of BWO, wherein substrate-induced epitaxial strain and symmetry have a negligible impact on ferroelectric switching.

The ability to grow high-quality BWO thin films, as demonstrated in this study, could unlock the full potential of BWO, paving the way for magnetoelectric and possible spintronics applications. Future studies could be aimed at investigating thickness scaling and the strain dependence of in-plane ferroelectricity. Ferroelectric hysteresis loop measurements using interdigitated electrodes would provide information on saturation/remnant polarization, which would complement spontaneous polarization values estimated using the STEM technique. A systematic study is also required to understand why the nominally charged head-to-head and tail-to-tail domain walls in the BWO film are insulating.

**MATERIALS AND METHODS**

**SAMPLE FABRICATION.** BWO films used for constructing the growth phase diagram and STEM measurements were grown on 2 unit cells thick SRO buffered (001)-oriented $SrTiO_3$ (STO) substrates. For the BWO film growth, a polycrystalline 10% Bi-rich BWO target was ablated using a KrF excimer laser ($\lambda$ = 248 nm). The laser fluence, spot size, and target-to-substrate distance were fixed at 1.3 J cm$^{-2}$, 2.2 mm$^2$, and 5 cm, respectively. After sample growth, the films were cooled to room temperature under growth pressure at a cooling rate of 30°C/min. For the BWO film used in the PFM measurement, the growth of 8 nm (21unit cells) thick SRO bottom electrode was followed by the growth of an 11-unit-cells-thick BWO film at $T_{sub}$ = 600°C and $P_{O_2}$ = 125 mTorr. The growth was monitored by the *in situ* reflection high energy electron diffraction technique (Figure S8). After film growth, the BWO film was annealed at 400°C under ambient oxygen pressure to remove excess oxygen vacancies and subsequently cooled to room temperature. Substrate temperatures were recorded by using an optical pyrometer facing the backside of the SiC substrate holder. Insertion of the SRO layer is an effective means to stabilize single-phase BWO film on the STO substrates, which is otherwise hindered by the diffusion of oxygen vacancies during the growth[24].

**STRUCTURE ANALYSIS.** XRD measurements were performed using a Bruker AXS D8 X-ray diffraction (XRD) machine equipped with a Cu K α-1 source ($\lambda$ = 1.5406 Å) and a Vantecline detector.

**STEM MEASUREMENTS.** The electron-transparent cross-sectional STEM specimen was prepared by focused ion beam milling (Helios 650; FEI/Thermo Fisher Scientific, Waltham, MA, USA) and further thinned by focused Ar-ion milling (NanoMill 1040; Fischione Instruments, Export, PA, USA). STEM images were acquired at room temperature using a spherical aberration-corrected STEM (JEM-ARM 200F; JEOL Ltd., Tokyo, Japan) equipped with a cold field-emission gun operated at an electron acceleration voltage of 200 kV. The semi-convergence angle of the electron probe was 24 mrad, and the collection inner semi-angle for HAADF was 80 mrad. The collection semi-angles for ABF were set to 12–24 mrad. The low angle annular dark field (LAADF) images (Figure S7) were recorded with a collection semi-angle of 30 mrad.

All HAADF and ABF STEM images in this paper were obtained by averaging 20 consecutive images (1,024 pixels × 1,024 pixels) captured with a 2-μs/pixel dwell time. LAADF images (Figure S8) were obtained by averaging 40 consecutive images (1,024 pixels × 1,024 pixels) captured with a 1-μs/pixel dwell time. Each image set was then averaged out by a rigid registration method for accurate measurement of atomic positions, using 2D Gaussian fitting with seven parameters and custom MATLAB scripts. W-cation displacement was calculated based on the center of the two-nearest Bi-atomic positions[25].

**PFM MEASUREMENT.** PFM was performed using an Asylum Research Cypher atomic force microscope under ambient conditions at room temperature. The PFM images were obtained by applying an alternating voltage of amplitude $V_{ac}$ = 700 mV through a conducting AFM tip (DDESP-FM-V2, Bruker).

## ASSOCIATED CONTENT

**Supporting Information.** The Supporting Information is available free of charge (.pdf). The Supporting Information contains the following.

Rocking curve analysis, reciprocal space mapping, RHEED data during in-situ growth, AFM images, and additional STEM images.

## AUTHOR INFORMATION

Corresponding Author

*Saikat Das (DAS.Saikat@nims.go.jp); ORCID id : orcid.org/0000-0002-3857-3096

Present Addresses

Research Center for Magnetic and Spintronic Materials, National Institute for Materials Science, 1-2-1 Sengen, Tsukuba 305-0047, Japan.

**Author Contributions**

[‡] These authors contributed equally to this work. S.D. conceived the project under the direction of T.W.N. J.J. and S.D. grew the samples. J. J. and J.K. performed the XRD measurements; J.M. carried out the STEM measurements and analysis under the supervision of M.K. S.D. and J.J. performed the PFM measurements. J.R.K. and W.P. assisted in the growth and ferroelectric characterization, respectively. S.D wrote the manuscript with the feedback from all the authors.

## ACKNOWLEDGMENTS

This work was supported by the Institute for Basic Science in Korea IBS-R009-D1. J. M. and M.K. acknowledges the support from National Center for Inter-University Research Facilities (NCIRF) in Seoul National University in Korea.

## NOTES

The authors declare no competing financial interest.

# Supporting Information

# Growth and atomically resolved polarization mapping of ferroelectric $Bi_2WO_6$ thin film

Jihwan Jeong[1,2,‡], Junsik Mun[3,1,‡], Saikat Das[1,2,†,*], Jinkwon Kim[1,2], Jeong Rae Kim[1,2], Wei Peng[1,2], Miyoung Kim[3,1], and Tae Won Noh[1,2]

[1] Center for Correlated Electron Systems, Institute for Basic Science (IBS), Seoul 08826, Korea.

[2] Department of Physics and Astronomy, Seoul National University, Seoul 08826, Korea.

[3] Department of Materials Science and Engineering and Research Institute of Advanced Materials Seoul National University Seoul 08826, Republic of Korea.

[‡] These authors equally contributed to this work.

* Corresponding author's Email: DAS.Saikat@nims.go.jp

[†] Present Addresses

Research Center for Magnetic and Spintronic Materials, National Institute for Materials Science, 1-2-1 Sengen, Tsukuba 305-0047, Japan.

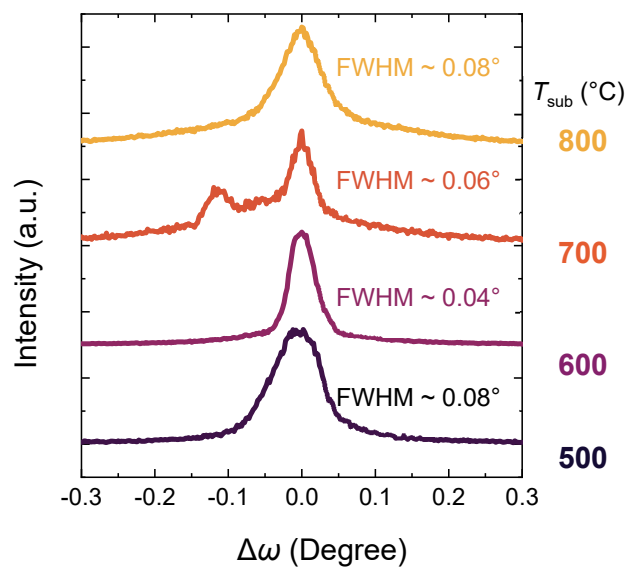

**Figure S1**. Rocking curves measured around the BWO (006) peak. The fullwidth at half maximum (FWHM) values are extracted by fitting the data to the pseudo-Voigt function. The films are grown at $P_{O_2}$ = 125 mTorr, by varying $T_{sub}$ from 500 °C–800 °C, and the corresponding XRD spectra are shown in Fig. 1b of the main text.

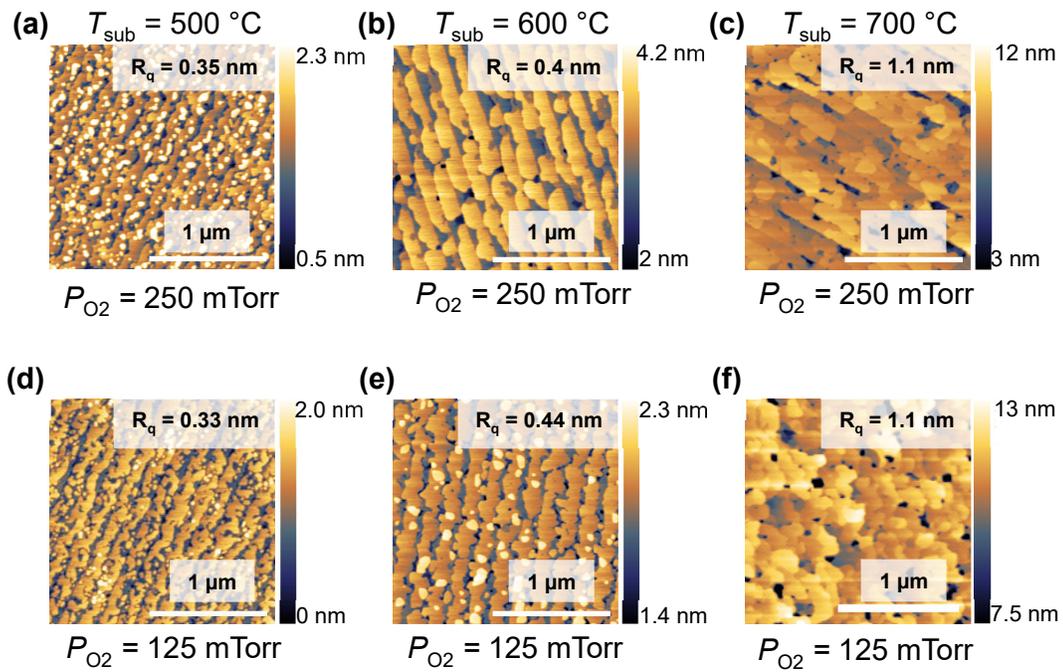

**Figure S2**. (a,b,c)AFM topography images of BWO thin films grown under oxygen partial pressure of 250 mTorr with the substrate temperature set to 500 °C (a), 600 °C (b), and 700°C (c). (d,e,f)AFM topography images of BWO thin films grown under oxygen partial pressure of 125 mTorr with the substrate temperature set to 500 °C (a), 600 °C (b), and 700°C (c). The root mean square roughness ($R_q$) is provided in the inset.

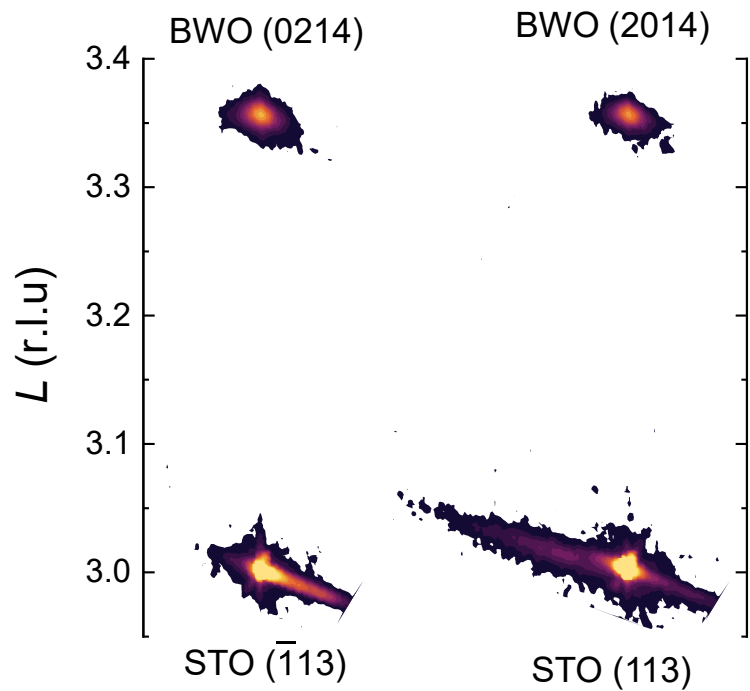

**Figure S3**. Reciprocal space mapping around the STO ($\bar{1}$13) and STO (113) Bragg reflexes. The BWO film used for this measurement was grown at $T_{sub}$ = 600 °C and $P_{O_2}$ = 125 mTorr, and the thickness is around 12 unit cells.

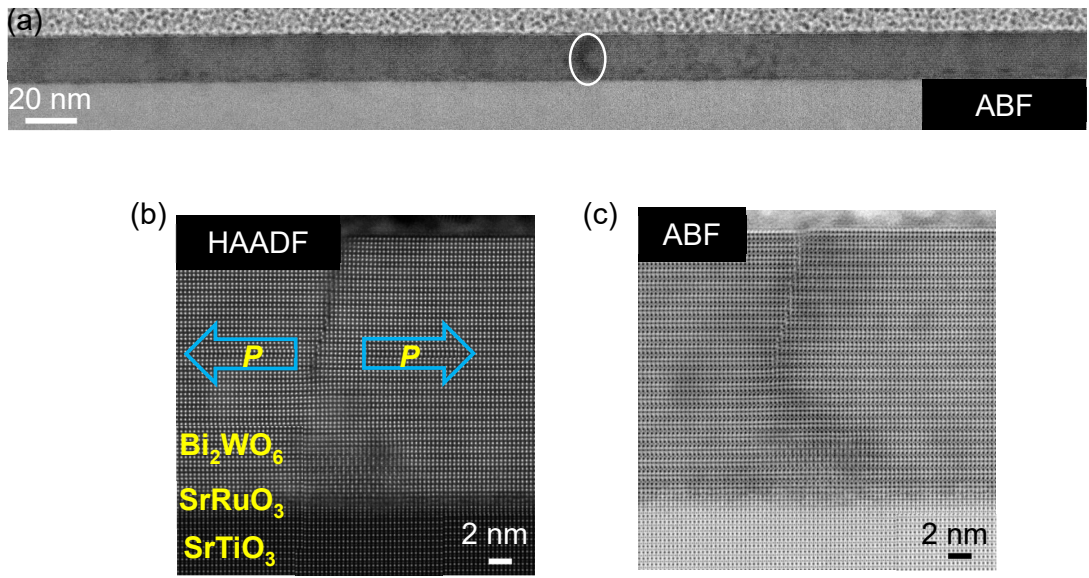

**Figure S4.** (a) Low magnification ABF-STEM image of the BWO film grown at $T_{sub}$ = 600 ºC and $P_{O_2}$ = 125 mTorr. This image shows a coherent film growth over a wide lateral scale and a defective region marked by the white ellipse. (b)-(c) Magnified HAADF (b) and ABF image (c) around the defective region in (a). Magnified images show the defective region consists of an out-of-phase boundary starting from a defect near the SRO/STO interface. Interestingly, the polar displacement of W atoms is opposite on either side of the out-of-phase boundary forming a tail-to-tail domain wall as shown by the arrows. The STEM images were obtained along the BWO [100] zone axis.

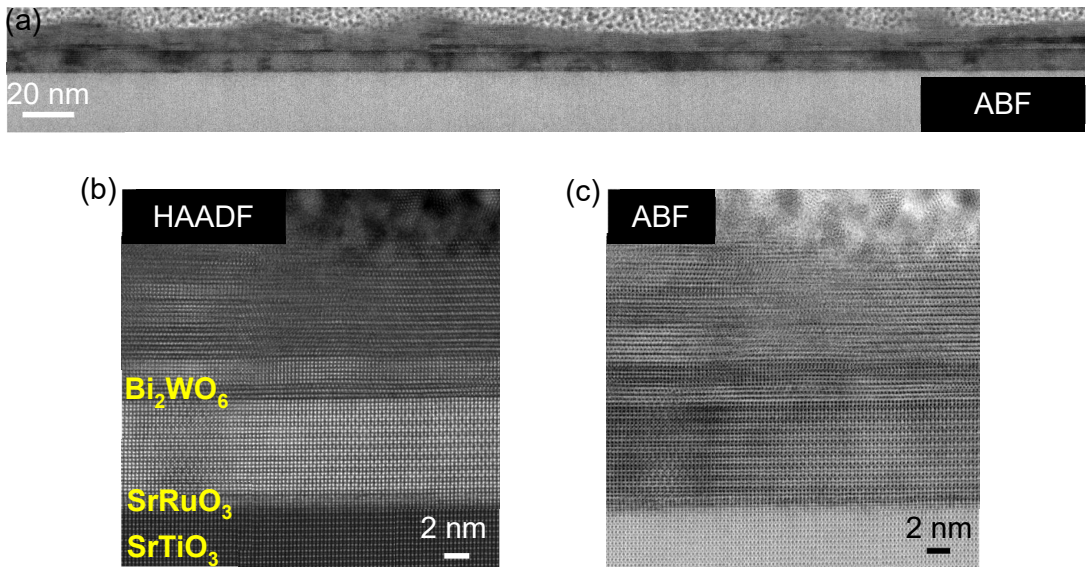

**Figure S5.** (a) Low magnification ABF-STEM image of the BWO film grown at $T_{sub}$ = 600 ºC and $P_{O_2}$ = 75 mTorr. Darker image contrast implies the presence of excessive defects in the film. by the white ellipse. (b)-(c) Magnified HAADF (b) and ABF image (c) showing Bi deficiency and presence of non-layered phase near the surface region of the film. The STEM images were obtained along the BWO [100] zone axis

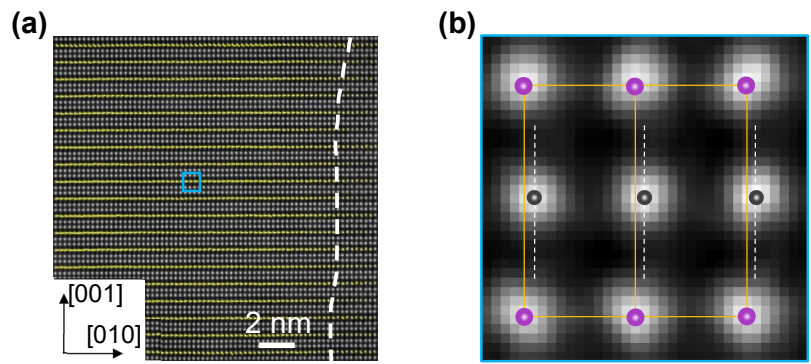

**Figure S6.** (a) HAADF-STEM image that was taken along the BWO [100] zone axis and it shows the polar displacement around the type-I/type-II domain wall. The dashed line marks the domain wall. (b) Magnified HAADF image showing the polar displacement is along the BWO [010] direction, opposite to that shown in Figure 3 of the main text.

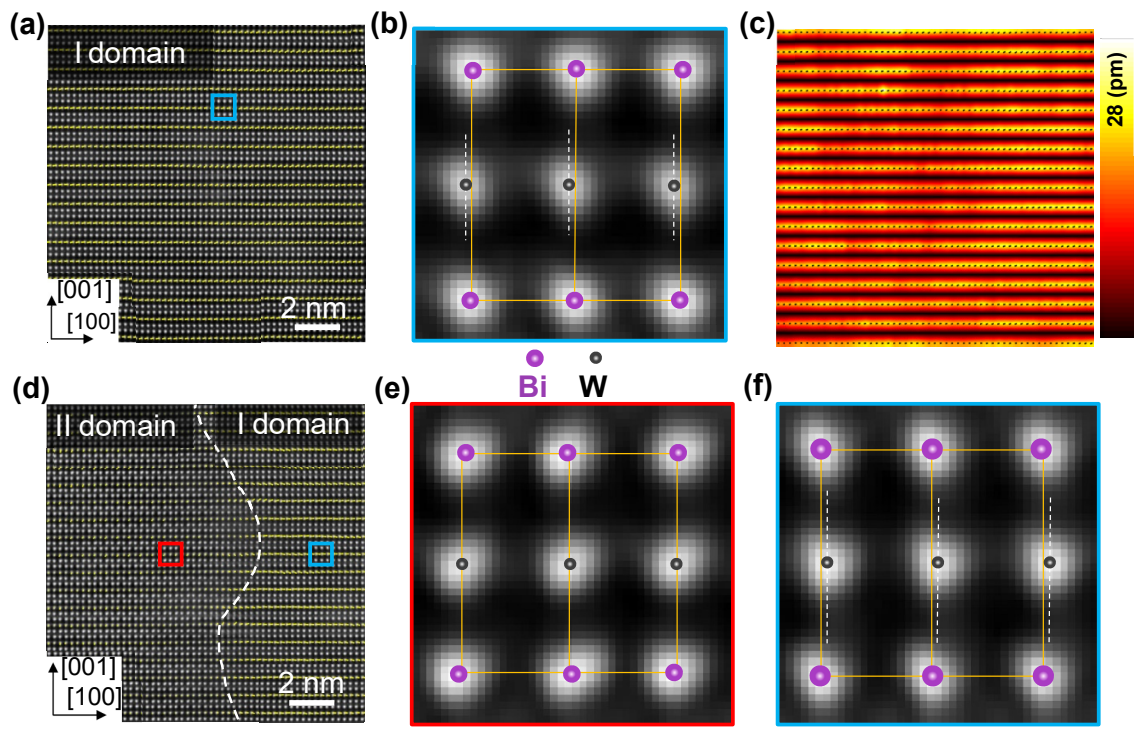

**Figure S7**. Characterization of polar displacement in using the low angle annular dark field (LAADF) STEM imaging. (a) High-resolution LAADF image of the type-I domain. The yellow arrows mark the direction of polar displacements. (b) Magnified image showing the polar displacement of the W atoms (black spheres) relative to the Bi atoms (magenta spheres) within the area marked by the cyan colored rectangle. (c) False color maps showing the layer-resolved magnitude of polar displacement. We estimated the average polarization value to be about $50 \pm 4$ μC cm$^{-2}$, consistent with the value estimated from the HAADF measurements. (d) High-resolution LAADF image across the 90° domain wall marked by the white dashed line. (e) Magnified image from the type-II domain (marked by red colored square in (d)), showing the centrosymmetric arrangement of W atoms against the Bi atoms. (f) Magnified image from the type-I domain (marked by cyan colored square in (d)), showing the polar displacement of W atoms against the Bi atoms. Note the direction of polar displacement in (f) is opposite to that in (b). The LAADF images in (a) and (d) were obtained along the BWO [010] zone axis.

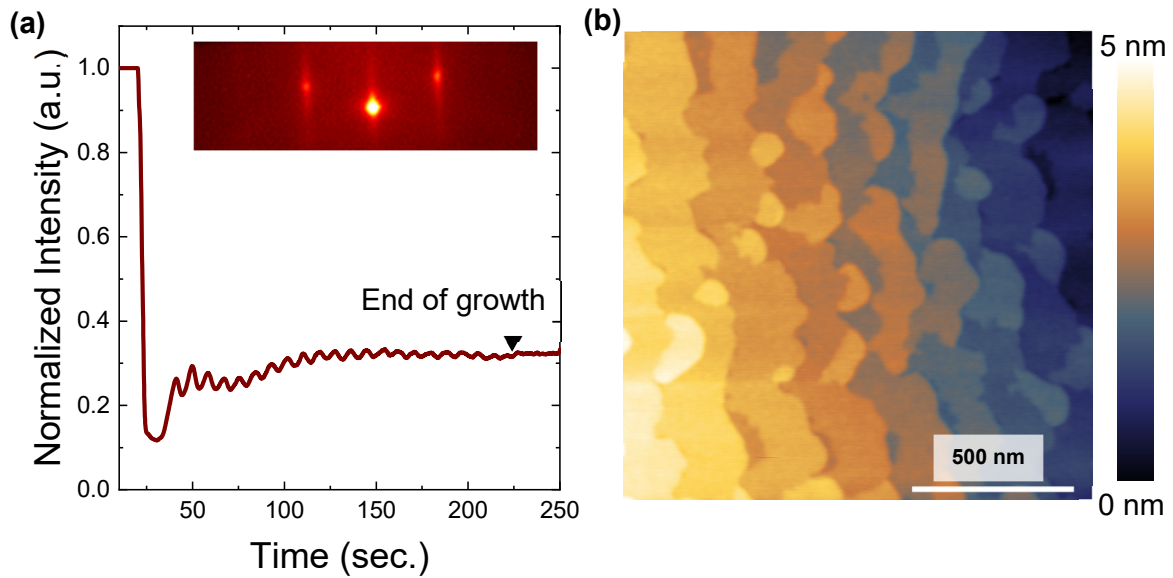

**Figure S8**. (a) *In situ* reflection high energy electron diffraction (RHEED) characterization profile during the growth of the 11unit cells thick BWO film that was used for the PFM measurement. The RHEED intensity show oscillations, suggesting layer-by-layer growth. Two complete oscillations correspond to the deposition of one structural unit cell of BWO. Inset shows the RHEED pattern at the end of the BWO growth. (b) The topography of the BWO film showing step-terrace structure; the step height ~ 0.4 nm, that is a quarter of the *c* axis lattice parameter.

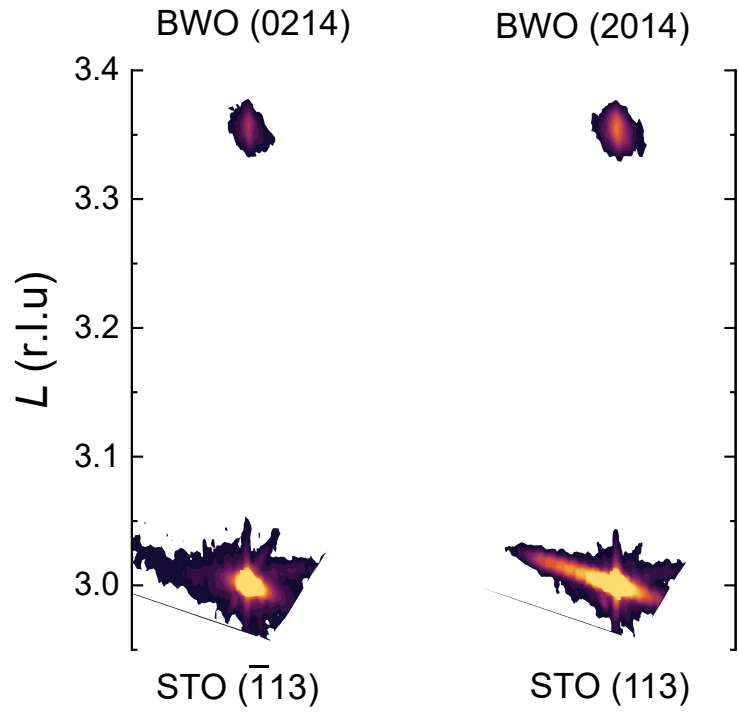

**Figure S9**. Reciprocal space mapping around the STO ($\bar{1}$13) and STO (113) Bragg reflexes. The BWO film used for this measurement was grown on 20 unit cells thick SRO layer and used for the PFM measurements in Figs. 4 and 5 of the main text. We estimated the in-plane lattice parameters of the BWO film are about a = 5.5223±0.0005 Å and b = 5.5174±0.00017 Å. These values are fairly comparable to the BWO film grown on 2 unit-cells thick SRO layer. This implies that the strain state is well preserved on the sample grown on the thicker SRO layer.

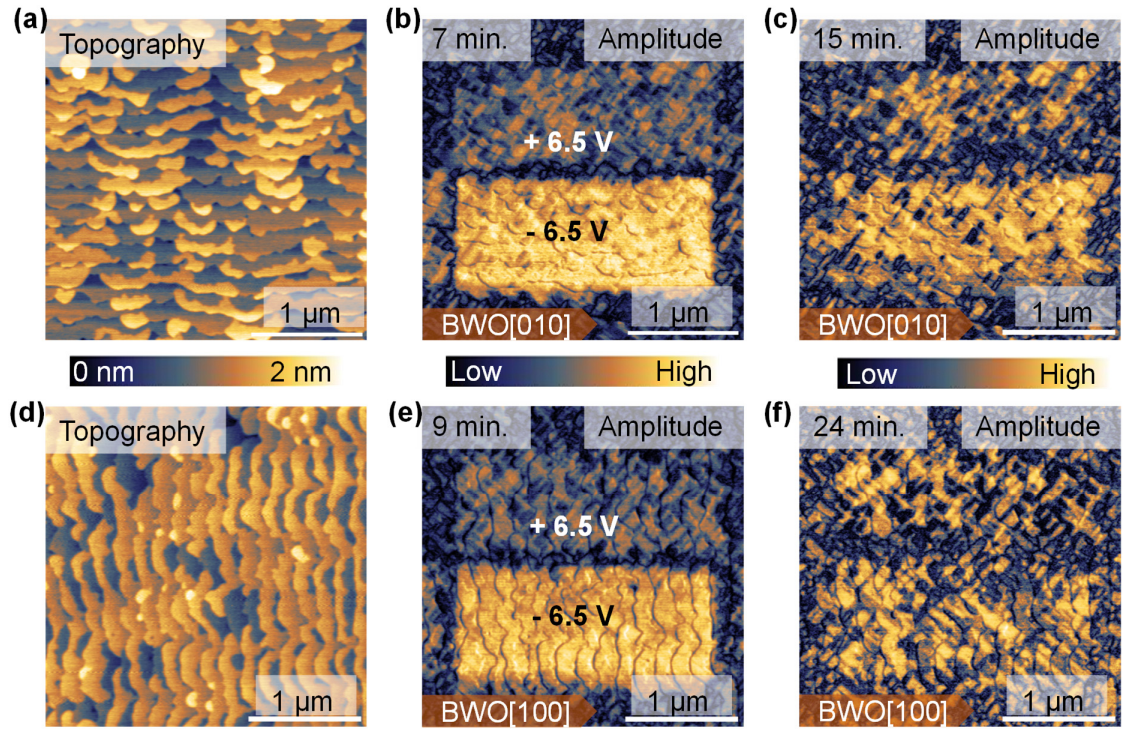

**Figure S10**. (a) AFM topography image. (b, c) PFM amplitude images after seven (a) and fifteen minutes from electrical poling. Images in (a)-(c) are obtained by aligning the cantilever axes parallel to the BWO [010] axis. (d) AFM topography image. (e, f) PFM amplitude images after nine (a) and twenty-four minutes from electrical poling. Images in (d)-(f) are obtained by aligning the cantilever axes parallel to the BWO [100] axis. Topography images in (a) and (d) correspond to PFM images shown in Figures 4 and 5 of the main text. Please note PFM amplitude images in (b) and (e) are the same as in Figs. 5 (a) and (c) of the main text. The amplitude contrast between positively and negatively poled regions diminishes over time. This suggests that during electrical poling, charge injection from the biased tip contributes to the PFM amplitude signal, which is more predominant for the negative bias case.
33

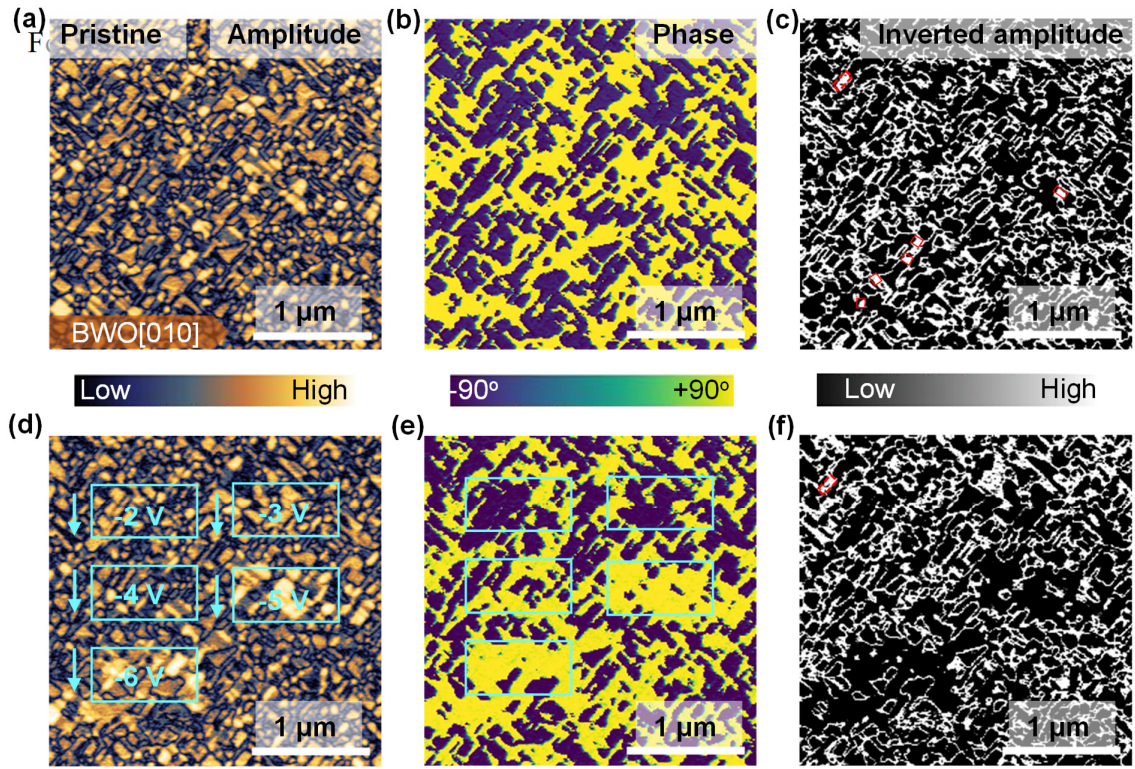

**Figure S11**. (a, b, c) PFM amplitude (a), phase (b), and the amplitude image on an inverted color scale (c) of pristine BWO film. Images are obtained by aligning the cantilever axes parallel to the BWO [010] axis. (d, e, f) PFM amplitude (d), phase (e), and the amplitude image on inverted color scale (f) after electrical poling the regions marked by the cyan colored rectangles in (d)-(e). The tip bias was varied from -2V to -6V, while the slow scan axis was parallel to the BWO [010], as marked by the vertical arrows in (d). The red colored solid rectangles mark the 90° DWs in the region poled with -5V and -6V. The red colored dashed rectangles in (c) and (f) mark the reference 90° DW to account for the sample drift before and after poling. This experiment shows that domain switching onsets for the applied bias ≥ - 4V and almost full switching for the applied bias of -5V and -6V. We observed an electrical breakdown for bias ≥ -7V. Although we could identify random patches of ferroelastic domains marked by solid rectangles in (c), within the narrow bias window below electrical breakdown, we could not observe partial switching, i.e., only ferroelectric switching and no ferroelastic switching.